\documentclass{article}[12pt]

\usepackage{graphics}
\usepackage{latexsym}  

\def\b{\beta}

\def\ve{\varepsilon}

\def\F{\Phi}
        
\def\k{\kappa}       
               
\def\L{\Lambda}

\def\id{\protect{{1 \kern-.28em {\rm l}}}}

\def\N{{\cal N}}
\def\F{{\cal F}}

\newcommand{\ti}[1]{\tilde{#1}}       
       
\newcommand{\sm}[1]{\mbox{\scriptsize #1}}        
       
\renewcommand{\@}[1]{\sqrt{#1}}       
\newcommand{\Tr}{{\mbox{Tr}}\,}       
\renewcommand{\le}[1]{\label{#1}\end{eqnarray}}        
\newcommand{\be}{\begin{equation}}
\newcommand{\ee}{\end{equation}}
\newcommand{\bea}{\begin{eqnarray}}
\newcommand{\eea}{\end{eqnarray}}
\newcommand{\nn}{\nonumber}

\newcommand{\eq}[1]{(\ref{#1})}        
\def\nn{\nonumber\\}

\def\ffract#1#2{\raise .35 em\hbox{$\scriptstyle#1$}\kern-.25em/       
\kern-.2em\lower .22 em \hbox{$\scriptstyle#2$}}

\def\half{{1\over2}\,}
\begin{document}

\large

\begin{titlepage}

\begin{flushright}
UCLA/02/TEP/40\\
{\tt hep-th/0212083}
\end{flushright}
\vskip0.1truecm
\begin{center}
\vskip 2truecm
%
%
%
%
%
{\Large\bf
Generalized Yukawa couplings and Matrix Models}
\vskip 1.5truecm
{\large\bf Iosif Bena${}^{\star}${}\footnote{
e-mail: {\tt iosif@physics.ucla.edu}},
{\large\bf Sebastian de Haro${}^{\star}${}\footnote{e-mail: {\tt sebas@physics.ucla.edu}}}
and Radu Roiban${}^{\dagger}$\footnote{
e-mail: {\tt radu@vulcan2.physics.ucsb.edu}}}\\
\vskip .5truecm
${}^{\star}$ {\it Department of Physics and Astronomy\\
University of California\\
Los Angeles, CA 90095}
\vskip 4truemm
${}^{\dagger}$ {\it Department of Physics\\
University of California\\
Santa Barbara, CA 93106} \\
\vskip 2truemm
\end{center}
\vskip 1truecm
\begin{center}
{\bf \large Abstract}
\end{center}
In this note we investigate $U(N)$ gauge theories with matter in the fundamental and adjoint 
representations of the gauge group, interacting via generalized Yukawa terms of the form $\Tr[Q 
\Phi^n {\tilde Q}]$. We find agreement between the matrix model and the gauge theory
descriptions of these theories. The analysis leads to a partial  description of the Higgs branch of 
the gauge theory. We argue that the transition between phases with different unbroken flavor symmetry 
groups is related to the appearance of cuts in the matrix model computation. 

\end{titlepage}


\section{Introduction}

Recently, Dijkgraaf and Vafa have 
proposed \cite{dv1,dv2,dv3} a method for computing perturbatively
the effective glueball superpotential of $\N=1$ theories with fields
transforming in the adjoint and 
bifundamental representations of the gauge group.
According to this proposal, the planar free energy of
the matrix model whose potential is the tree-level superpotential of the
${\cal N}=1$ theory yields the effective superpotential of this theory.

When fields transforming in the fundamental representation of the gauge
group (quarks) are present, one only needs to include the planar free
energy coming from diagrams with one quark boundary \cite{ACFH,BERO}. More
explicitly, the gauge theory effective superpotential is proposed to be
\be
W_{\sm{eff}}(S,\Lambda)=N_cS(1-\ln{S\over \Lambda^3}) + 
N_c{\partial \F_{\chi=2}\over \partial S}
+N_f \F_{\chi=1}~~.
\label{prescription}
\ee

This prescription was successfully used to compare matrix model
predictions with known gauge theory results for theories with massive 
and massless flavors, with ${\cal N}=1$ and ${\cal N}=2$ supersymmetry
\cite{ACFH}-\cite{ohta}.

For theories with fields transforming in the adjoint representation of the gauge group,
proofs that planar graphs are the only ones which contribute to the
matrix effective superpotential were presented  in \cite{grisaru} (based on 
the analysis of superspace Feynman diagrams) and \cite{witten} (based on 
holomorphy and symmetries). The latter arguments were extended in \cite{rrr}
to the case of theories with fields transforming in the fundamental representation of the 
gauge group and it was shown that only planar diagrams with one (appropriately 
generalized) quark boundary contribute to the gauge theory effective superpotential.
Other interesting related work has appeared in
\cite{MCGR}-\cite{new}.

The correspondence between gauge theories and matrix models has been
pushed very far for superpotentials depending only on the adjoint fields.
However, these checks have only been performed in the simplest cases of 
theories with fields transforming in the fundamental representation of the 
gauge group. While it seems reasonable that the arguments of
\cite{witten} and \cite{rrr} generalize to generic superpotentials,
it would be interesting to perform some explicit checks, along the lines of 
\cite{ACFH} and \cite{BERO} .

In this paper we work out the details of the matrix model and the gauge
theory for the $\ti Q\Phi^nQ$ coupling. With these results as a starting point, we
then outline how a polynomial of generalized Yukawa couplings $\ti Q P(\Phi)Q$ can be
analyzed. We find complete agreement between the matrix model with
one boundary and the gauge theory. The Coulomb branch of such theories was discussed
in detail in \cite{KAPU}. However, the Higgs branch seems largely unexplored. 
The matrix model computations suggest a simple way for analyzing it.

In the next section we use the matrix model to compute the
effective superpotential
of this theory. Since the adjoint field does not interact with itself,
this superpotential is just the sum of a Veneziano-Yankielowicz
piece (coming from the dynamics of the gauge field) and the free energy
given by diagrams with one quark boundary.

We find that the sum of these diagrams gives a free energy identical to
that of a theory containing an adjoint and $n$ quarks 
with regular Yukawa couplings $g_i \tilde Q_i \Phi Q_i $, 
where the coupling constants $g_i$ are proportional to the $n$ roots 
of the unity. We then show that one can go 
from the second theory to the first by simply 
integrating out certain combinations of the $n$ quarks until only one 
quark and the adjoint are left. 

We then discuss the gauge theory origin of the matrix model results. 
The nonperturbative contribution 
to the effective superpotential of the theory with a $\tilde Q \Phi^n Q$ 
coupling is hard to obtain by symmetry and holomorphy arguments. One might
hope that integrating $\Phi$ out might make things better, since only quarks 
will be left and the nonperturbative contribution to the superpotential
would be of Affleck-Dine-Seiberg type \cite{ADS}. Nevertheless, after integrating out
$\Phi$, one is left with a ``tree level'' superpotential which contains 
the coupling constant to a negative power. Since this term does not have a
well defined limit as $g \rightarrow 0$, one can no longer argue that this
term cannot mix with the Affleck-Dine-Seiberg contribution. Therefore, one
expects nonperturbatively generated  terms which contain combinations of
$\Lambda$ and $g$. These nonperturbative terms cannot be easily
found using analyticity and charge conservation.

It appears therefore that in order to find the nonperturbatively generated
contribution to the superpotential one has to find a theory (related to
the theory of interest by integrating in and integrating out) where the
nonperturbative contribution has a simple form. Fortunately, as the matrix
result hints also, the theory with Yukawa coupling 
$ \Tr[Q \Phi^n {\tilde Q}] $ can be obtained by integrating out $n-1$ nontrivial
linear combinations of quarks in a theory with an adjoint, $n$ quarks, and 
Yukawa couplings $ \sum_{l=1}^n e^{2 \pi
i l/n }  \Tr[Q_l \Phi {\tilde Q}_l] $. 
At this stage one can integrate out the adjoint field and find a theory with
$n$ quarks and interactions of the form $ \Tr[(\sum_l g_l Q_l {\tilde Q}_l)^2] $. For
this theory one can use the usual holomorphy and charge conservation
arguments \cite{seiberg} to show that the only nonperturbatively generated
superpotential is the Affleck-Dine-Seiberg one. 

Once we have the full gauge theory effective potential it is not hard to
integrate out all the fields and relate the resulting effective
superpotential  $W_{\sm{eff}}(\Lambda,m_i,g_i)$ with the matrix model
computation. The Higgs branch of the original theory can also be analyzed.

We emphasize that his effective superpotential is the same,
regardless of the order in which one integrates out the fields\footnote{We are grateful to 
Eric D'Hoker for pointing this out to us.}, and 
thus regardless of the hierarchy of the $m_i$. This is quite obvious from the matrix model
perspective: by summing all (planar) Feynman diagrams one obtains the same function of mass 
parameters, regardless of their hierarchy. 

One can also see this in the gauge theory. The physical mass of a field
depends both on its superpotential mass parameter $m$, and on the K\"ahler
potential. Since by adjusting the latter any hierarchy can be achieved
(regardless of the magnitudes of the $m$'s), and since this adjustment
does not affect the superpotential, it follows that  the final result is
independent on the mass hierarchy and on the order one integrates fields
out. 

Therefore, the effective superpotential $W_{\sm{eff}}(\Lambda,m_i,g_i)$ one
finds after integrating out all the fields is the same if the theories we
start from can be related to each other by integrating in or integrating
out. In the case of the theory we discuss, this effective superpotential
is most easily obtained by considering the related theory with $n$ quarks and
simple Yukawa interactions, finding its nonperturbatively generated
superpotential, and integrating out all the quarks. This computation
appears in the last section of this note. 

{\bf Note Added:} When this work was near completion we received the 
preprint \cite{ohta} which, while having a different focus, overlaps with 
the technical details of our work.

\section{Matrix model}

As we recalled in the Introduction, the part of the matrix model free energy $\F(S)$ 
which is related to gauge theory via extensions of the DV prescription 
is computed by summing all the planar Feynman diagrams with one
quark boundary. If the quark-adjoint interaction is of the form $\tilde Q_i \Phi
Q_i $, the necessary combinatorics of the Feynman diagrams was described in
\cite{ACFH,brezin}:  a diagram with $2k$ vertices comes with a factor of
${1\over (2 k) !}$ from the exponential; then, the different ways of contracting the
quarks produce a factor of $(2k-1)!$; the number of ways of connecting, in a 
planar manner, the boundary points with adjoint propagators give 
rise to a factor  ${(2k) ! \over (k+1)! k !}$. Since each diagram contains 
$2 k$ quark propagators, $k$ adjoint propagators, and $2k$ vertices, it is 
multiplied by $\left({g^2 \over M_{\Phi} m^2_{Q}}\right)^k$. Finally, the external 
flavor loop gives a
factor of  $N_f$, while the $k+1$ color loops give a factor of $S^{k+1}$,
where $S$ is the 't Hooft coupling of the matrix model, and becomes
identified under the correspondence with the gauge theory glueball
superfield.

Thus, the free energy contribution of these diagrams is \cite{ACFH}
\be
\F^{n=1}_{\chi=1}= - N_f \sum_{k=1}^{\infty} { (2k-1) ! \over (k+1)! k !}\,
S^{k+1} \left({g^2 \over M_{\Phi} m^2_{Q}}\right)^k
\ee

In the case of diagrams with a $g_n \tilde Q_i \Phi^n Q_i $ interaction,
it is necessary to distinguish between the case of odd and even $n$. If $n$ is
odd, only diagrams with even numbers of insertions contribute. The
combinatorial factors coming from the quarks are unchanged. Nevertheless
since now there are $ n k$ $\Phi$ lines, there will be ${(2 n k)! \over (nk +1) ! (nk)!}$ 
different ways of connecting them, and the overall power
of $S$ will be $nk+1$. Thus
\be
\F^{\sm{odd}}_{\chi=1}= - n N_f \sum_{k=1}^{\infty} { (2nk-1) ! \over
(nk+1)! (nk) !}\, S^{nk+1} \left({g_n^2 \over M^{n} _{\Phi}m^2_{Q}}\right)^k
\label{Fodd}
\ee

If $n=2 p$ is even, diagrams with any number of insertions
contribute. The free energy is
 \be
\F^{\sm{even}}_{\chi=1}= - 2 p N_f \sum_{k=1}^{\infty} { (2pk-1) ! \over
(pk+1)! (pk) !}\, S^{pk+1} \left({g_n \over M_{\Phi}^{p}
m_{Q}}\right)^k
\ee

One might have also expected extra factors of $n!$ coming from the
different orderings of the $\Phi$'s originating from one vertex. However,
since the $\Phi$'s are matrices, one cannot interchange them at a vertex
because this would make the diagrams nonplanar. Thus, the diagrams give the
same answers as if the $n$ $\Phi$'s originating at one interaction vertex
were separated by tiny propagators of some auxiliary quarks. This is
depicted in Figure 1, and is a hint toward the equivalence of our theory
to a theory with $n$ quarks and simple Yukawa couplings, equivalence which
will be discussed in the next section. 
\begin{center}
\begin{figure}[ht]
\centerline{\scalebox{.57}{\includegraphics{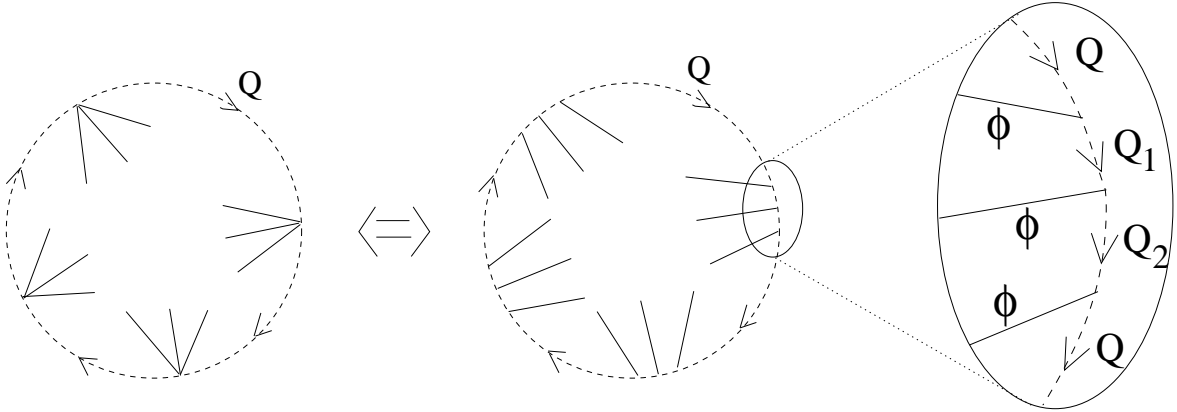}}}
\caption{Equivalence between Yukawa and generalized Yukawa couplings.}
\end{figure}
\end{center}
\vspace{-1cm}

If one introduces the new variable $p$, such that $p \kappa = n$ where 
\bea
\k&=&2 ~~~~~~~~~~~~\mbox{if $n$ is even},\nn
\k&=&1~~~~~~~~~~~~\mbox{if $n$ is odd}~,
\eea
it is not hard to see that the free energy for both even or odd $n$ is given by:
\be
{\cal F}_{\chi=1}=-\k p N_f{\cal K}_p,
\label{f}
\ee
where ${\cal K}_p$  is the value of the sum in (\ref{Fodd}). 

The radius of convergence of the sum ${\cal K}_p$ can be easily found to be 
$S_c={1\over 4 \alpha}$. For $S<S_c$ the sum is: 
\be
{\cal K}_p=
{1\over p}\sum_{l=0}^{p-1}S\left[{1\over 2}+{1\over 4 (-)^{2{l\over p}} \alpha S}
\left[\sqrt{1-4(-)^{2{l\over p}}\alpha S}-1\right]
-\ln{1\over 2}(1+\sqrt{1-4(-)^{2{l\over p}}\alpha S})
\right]\nonumber
\label{Kp}\ee
and $\alpha^{n} = \left({g_n^2 \over M_{\Phi}^{n} m^2_{Q}}\right)$. 

For  $S\ge S_c$ the sum is divergent, and one must find its value by
analytical continuation \cite{ohta}. Since (\ref{Kp}) contains square roots, 
one expects branch cuts in the complex $S$ plane starting 
from the points where the square roots become zero and ending at infinity. 

Moreover, since we have $n$ square roots, each
comes with a choice of branch. Therefore, for $S$ outside the radius of
convergence, the sum (\ref{Kp}) has $2^{n}$ values, depending on the choice of
branch for each square root. Furthermore, since the sum  (\ref{Kp}) applies separately 
for each of the flavors, there will be a total of $2^{nN_f}$ branches for the free energy. The result is:  
\be
{\cal K}_p(\epsilon) ={1\over p}\sum_{l=0}^{p-1}S\left[{1\over 2} +{1\over 4 (-)^{2{l\over p}} \alpha S}
\left[\epsilon_{l,f} \sqrt{1-4(-)^{2{l\over p}}\alpha S}-1\right]
-\ln{1\over 2}\left(1+\epsilon_{l,f} \sqrt{1-4(-)^{2{l\over p}}\alpha S}\right)
\right]\nonumber
\label{Kp_a}\ee
with $\epsilon_{l,f}=\pm 1$ with $l=1,\dots,n$ and $f=1,\dots, N_f$ .

For the example discussed in  \cite{ohta},  the choice of branch in the matrix integral
was matched in gauge theory with the choice of roots of a certain the second order equations. 
However, one can choose the parameters of the theory such that
all relevant values of $S$ lie inside the radius of convergence. We limit ourselves to
showing agreement in this regime. We will show that the convergence of (\ref{Kp})
for $S<S_c$ has a precise meaning in gauge theory.
The branch structure and the corresponding phase structure of the theory \cite{ohta} 
follows from the careful analytical continuation of our results. 

To evaluate the superpotential (\ref{prescription}) at its critical point we begin
by constructing a single formula which covers both cases $\kappa=1$ and $\kappa=2$.
Since the set of odd roots of unity goes to itself when squared, it is not hard to see
that the free energy can be expressed solely in terms of $n$:
\be
{\cal F}_{\chi=1}=- N_f \sum_{l=0}^{n-1}S\left[{1\over 2}+{1\over 4 e^{2 \pi i{2l\over n}} 
\alpha S}
\left[\sqrt{1-4e^{2 \pi i {2 l\over n}}\alpha S}-1\right]
-\ln{1\over 2}(1+\sqrt{1-4 e^{2 \pi i{2 l\over n}}\alpha S})
\right]
\label{f_b}
\ee
Since all roots of order $n$ of unity appear in the above expression, it is clear that
all phases in the definition of $\alpha$ below equation (\ref{Kp}) are equivalent.

It is quite easy to see that each term in the sum in equation (\ref{f_b}) reproduces the 
1-boundary free energy of a theory with a single quark and a regular Yukawa coupling. 
The ratios of the coupling constants of these theories are $n$-th roots of unity. In the next section 
we will show how this comes about in the gauge theory.

Using  (\ref{f_b}), the critical points of the effective superpotential (\ref{prescription}) 
are given by the solution of the equation:
\be
{N_c\over N_f}\ln{S\over\Lambda^3}=
\sum_{l=0}^{n-1}\ln{1\over 2}(1+\sqrt{1-4e^{2\pi i{2l\over n}}\alpha\, S})=
\ln\prod_{l=0}^{n-1}{1\over 2}(1+\sqrt{1-4e^{2\pi i{2l\over n}}\alpha\, S})~~,
\ee
which can be trivially transformed into:
\bea
y^{N_c\over N_f}= \prod_{l=1}^n{1\over 2} (1+\sqrt{1-4e^{2\pi i{2 l\over n}} \beta \,y})~~,
\label{crMM}
\eea
with $y={S\over \Lambda^3}$ and $\beta=\alpha\Lambda^3 $.

Then, the values of the superpotential at its critical points are given by
\be
W|_{\sm{crit}}= \Lambda^3 \left[N_c-{n\over2}\, N_f + N_f\sum_{l=1}^n  
{1\over 1+\sqrt{1-4e^{2\pi i{2l\over n}}\beta\, y}}\right]\,y~~
\label{Wmatrix}
\ee
where we have to replace $y$ by a solution of (\ref{crMM}). This is the result that 
we will compare with the gauge theory predictions. We stress that 
this equation is valid only in the limit of small $\beta$, i.e. far from branch points of
the series (\ref{Kp}). 

For generic  $\beta$ the critical values of the superpotential are given by
\be
W|_{\sm{crit}}= \Lambda^3 \left[N_c-{n\over2}\, N_f + \sum_{f=1}^{N_f}\sum_{l=1}^n  
{1\over 1+\epsilon_{i,f}\sqrt{1-4e^{2\pi i{2l\over n}}\beta\, y}}\right]\,y~~,
\label{Wmatrix1}
\ee
where $y$ is given by an equation similar to (\ref{crMM}), except that the square roots are dressed 
with $\epsilon$ coefficients.

\section{Gauge theory}

Let us begin the gauge theory analysis by showing that the generalized Yukawa
coupling $\Tr[Q\Phi^n\ti Q]$ is equivalent to a set of $n$ ordinary $\Tr[Q\Phi\ti Q]$ 
terms whose strenghs differ by $n$-th roots of unity.  To this end we notice that,
by starting with the superpotential
\be
W_{\sm{tree}}={1\over 2} M\,\Tr \Phi^2 +m\sum_{i=1}^{n}\Tr[Q_i{\tilde Q}_i] +
g_0\sum_{i=1}^n \Tr [Q_i\Phi {\tilde  Q}_{i+1}]~~~~{\rm with}~
~{\tilde  Q}_{n+1}\equiv {\tilde  Q}_{1}
\label{off}
\ee
and integrating out $Q_i$ and ${\tilde Q}_i$ for all $i=2,\dots n$ we recover
\be
W_{\sm{tree}}={1\over 2} M\,\Tr \Phi^2 +m \Tr[Q{\tilde Q}] +g \Tr [Q\Phi^n {\tilde  Q}]
\ee
provided that $m^{n-1}g_0^n=g$. 

Equation (\ref{off}) can be rewritten as
\be
W_{\sm{tree}}={1\over 2} M\,\Tr \Phi^2 +m \sum_i \Tr[Q_{i}{\tilde Q}_i] +
m^{(n-1)/n}g_0 \sum_{i,j=1}^n\Tr [Q_i\Phi {\tilde  Q}_j]{\cal P}_{ij} 
\label{off1}
\ee
with the matrix ${\cal P}$ being given by
\be
{\cal P}=\pmatrix {0 & 0 & 0 &\dots & 0 & 1\cr
                   1 & 0 & 0 &\dots & 0 & 0\cr
                   0 & 1 & 0 &\dots & 0 & 0\cr
                   . & . & . &\dots & . & .\cr
                   . & . & . &\dots & . & .\cr 
                   0 & 0 & 0 &\dots & 1 & 0\cr}~~.
\ee
Since we chose the mass matrix of the $nN_f$ quarks to be proportional to 
the identity matrix, it is clear that the Yukawa and the mass terms can be
simultaneously diagonalized. Noticing that eigenvalues of  ${\cal P}$ are 
given by the roots of unity, we can immediately rewrite (\ref{off1}) as
\be
W_{\sm{tree}}={1\over 2} M\,\Tr \Phi^2 +m \sum_i \Tr[Q_{i}{\tilde Q}_i] +
\sum_{l=1}^n \omega_lg_0\Tr [Q_l\Phi {\tilde  Q}_l]~~~~{\rm with}~~
\omega_l=e^{2\pi i {l\over n}}~~.
\label{diag}
\ee
We have thus shown that the coupling $\Tr[Q\Phi^n{\tilde Q}]$ is equivalent to a set
of diagonal Yukawa couplings whose strengths differ by roots of unity. 

Unlike equation (\ref{off}), the superpotential (\ref{diag}) is invariant under global $SU(N_f)^{\otimes 
n}$
transformations. In constructing the Affleck-Dine-Seiberg superpotential it is useful to think of
(\ref{diag}) as a particular case of 
\be
W_{\sm{tree}}=\half M\Tr\Phi^2 +m\,\Tr[{\cal Q}\ti{\cal Q}] + \Tr[G{\cal Q} \Phi\ti{\cal Q}]~~,
\ee
where now ${\cal Q}$ are in $SU(nN_f)$ and $G$ is a diagonal matrix. The $SU(nN_f)$ invariance if 
broken either by having different coupling constants  or by having 
different masses for the $n$ sets of $N_f$ quarks.

Since we are interested in comparing this gauge theory to the matrix model of the previous section, we 
first integrate 
out the adjoint  fields. The effective superpotential is given by:
\be
W_{\sm{tree}}=m\Tr[{\cal Q}{\tilde {\cal Q}}] -{1\over2M}\Tr[G{\cal Q}\ti{\cal Q}G{\cal Q}\ti{\cal Q}]~~.
\ee
We will later identify the coupling constant matrix with the one given by  equation (\ref{diag}), 
but we will derive the formulae for an arbitrary diagonal $G={\rm diag}(g_1,\dots,g_n)$.

To this tree level superpotential we have to add the nonperturbative contributions.
It is not hard to see that they are given by the ADS superpotential for ${\cal Q}$ and ${\tilde {\cal 
Q}}$.
Indeed, in the limit of vanishing $m$ and $G$ this is the only possible term. Demanding analyticity 
as well as preservation of the symmetries leads to the conclusion that no corrections are possible.

Next we want to integrate out all quarks.  The easiest way to find the result is to rewrite the above 
superpotential in terms of mesons and notice that, since the mass matrix as well as $G$ are diagonal, 
all off-diagonal components of the mesons are constrained to vanish. Writing the remaining components of 
the 
meson field as $X_{ii}=Q_i{\tilde Q}_{i}=x_i \id_{N_f}$ the remaining effective superpotential is:
\bea
W_{\sm{eff}}=N_f\,m\sum_{i=1}^n x_i - N_f\sum_{i=1}^n a_i x^2_i +  
(N_c-nN_f)\left[{\Lambda^{3N_c-nN_f}\over \prod_{i=1}^n x^{N_f}_i}\right]^{1\over N_c-nN_f},
\eea
where $a_i={g_i^2\over2M}$. Minimizing this superpotential gives
\be
mx_i - 2 a_i x_i^2 - \Lambda^{3N_c-nN_f\over N_c-nN_f} \prod_{i=1}^n x_i^{-{N_f\over N_c-nN_f}}=0
\ee
To compare with the matrix model, it is useful to make the following change of variables:
\be
y=\prod_{j=1}^n y_j^{-{N_f\over N_c-nN_f}},
\ee
in terms of which the equations of motion can be rewritten as:
\be
y_i - \beta_i y^2_i - \prod_{j=1}^ny_j^{-{N_f\over N_c-nN_f}}=0
\label{fineq}
\ee
where we defined
\be
\beta_i={2a_i\over m}
\left[{\Lambda^{3N_c-nN_f\over N_c-nN_f}\over m}\right]^{N_c-nN_f\over N_c}\!\!\!\!\!\!\! ,
\ee
and we recall that $i=1,\dots, n$ labels the different types of quarks. This gives a  
system of $n$ coupled non-linear equations. To proceed it is helpful to 
introduce a ``radial'' variable:
\be
y=\prod_{j=1}^ny_j^{-{N_f\over N_c-nN_f}}.
\ee
The equation of motion becomes
\be
y_i - \beta_i y^2_i - y=0~~.
\label{eom}
\ee
We can solve for each individual $y_i$ in terms of the couplings and the radial variable:
\be
y_i=-{1\over 2\beta_i}\left[-1+\ve_i \sqrt{1-4\beta_iy}\right]~~,
\label{yi}
\ee
where $\ve_i=\pm1$. Each choice of $\ve$-s one finds an equation for $y$:
\be
y^{{N_c\over N_f}-n}=\prod_{i=1}^n {-2\beta_i\over -1+\ve_i \sqrt{1-4\beta_iy}}=
\prod_{i=1}^n 2\beta_i {1+\ve_i \sqrt{1-4\beta_iy}\over 4\beta_iy}~~.
\ee 
Bringing factors of $y$ together, we get the following equation:
\be
y^{N_c\over N_f}=\prod_{i=1}^n {1\over 2} {(1+\ve_i \sqrt{1-4\beta_iy})}
\label{yNcNf}
\ee
This algebraic equation can be solved numerically for various values of $n$ and the couplings. Once  
one  has a solution of this equation, one can obtain the $y_i$'s from equation \eq{yi} for each of 
the $2^n$ choices of $\epsilon_i$.

We next compute the effective superpotential at the minimum:
\bea
W|_{\sm{crit}}&=&{1\over 2}N_f\,m\sum_{i=1}^n x_i + {1\over 2} (2N_c-nN_f) \left[{\Lambda^{3N_c-nN_f}  
\over \prod_{i=1}^n x^{N_f}_i}\right]^{1\over N_c-nN_f}\nn
&=&m^{{nN_f\over N_c}} \Lambda^{3N_c-nN_f\over N_c } \left[{1\over 2}N_f  \sum_{i=1}^n y_i + {1\over  
2}  
(2N_c-nN_f)\prod_{i=1}^n y_i^{-{N_f\over N_c-nN_f}}\right].
\eea
In fact, the superpotential can be written in terms of the variable $y$ alone by using equation  
\eq{yi}
\bea
W|_{\sm{crit}}&=&\Lambda_0^3 \left[{1\over 2}(2N_c-nN_f)\,y +{1\over 2}N_f \sum_{i=1}^n{-1\over  
2\beta_i} \left(-1 +\ve_i\sqrt{1-4\beta_iy}\right)\right]\nonumber\\
&=&\Lambda_0^3
\left[N_c-{n\over2}\,N_f +N_f  \sum_{i=1}^n {1\over 1+\ve_i \sqrt{1-4\beta_iy}}\right]\,y
\label{Wcrit}\eea
where the scale $\L_0$ is defined by
\be
\Lambda_0^3=m^{{nN_f\over N_c}}\Lambda^{3N_c-nN_f\over N_c}
\ee
which is the correct relation between scales when all $nN_f$ quarks are integrated out.

So far the discussion was for general $\b_l$. Now, when all of the couplings $\b_l$, $l=1,\ldots,n$,  
are different, the number of solutions to the system \eq{eom}  is $2^n$  
times the number of solutions of equation \eq{eom}. If we chose a more 
general form of the meson, $X_{ii}={\rm diag}(x_i^1,\dots,x_i^{N_f})$, the number of different vacua would 
increase to 
$2^{nN_f}$ times the number of solutions of the apropriately modified equation (\ref{yNcNf}). This 
matches the 
total number of branch cuts in the matrix model computation. Furthermore, it is easy to see that 
specializing
the coupling constants $g_l$ to the ones implied by the equation (\ref{diag}) we find
\be
a_l=e^{2\pi i{2l\over n}}\,a=e^{2\pi i{2l\over n}}\,{g_0^2\over m^{2/n}}{m^2\over 2 M^2}~~~~\Rightarrow
~~~~
\beta_l=e^{2\pi i{2l\over n}}\beta~~.
\label{roots}
\ee

Therefore, the gauge theory and the matrix model results match in the region where the series (\ref{Kp})
develops branch cuts. However, since for $y\beta<{1\over 4}$ the series leading to (\ref{Kp_a}) is 
convergent, all $\epsilon_{i,f}$ coefficients are fixed to unity, 
while in gauge theory the choice of signs $\epsilon_{i,f}$ seems to persist for all values of $y$.

To understand the solution of this apparent discrepancy we should note that the
appearance of different branch points leads to a spontaneous breaking of the
$U(N_f)$ flavor symmetry in a theory  with $N_f$ identical quarks. In the regions of parameters 
where both $\epsilon_{i,f}=+1$ and $\epsilon_{i,f}=-1$ are allowed, there exist vacua with broken flavor 
symmetry.
However, one expects that, as the coupling constant is reduced, the flavor symmetry will be restored. 
Indeed, by 
taking the small $\beta_i$ limit on equation (\ref{yi}) we find that, for any $i$, the solution 
corresponding to $\epsilon_i=-1$ moves off to infinity while the one corresponding to $\epsilon_{i,f}=+1$ 
remains at 
finite distance. Thus, in the small coupling limit, only the choice $\epsilon_i=+1$ is allowed. 

This has definite meaning in the matrix model computation. In the small coupling limit the radius of 
convergence of
the series (\ref{Kp}) becomes very large. Therefore, all branch points move off to 
infinity and there remains a unique choice for the free energy ${\cal F}_{\chi=1}$.

We are therefore led to interpret the appearance of branch cuts in the series (\ref{Kp}) as the matrix 
model version of 
transitions between domains of the Higgs branch with different flavor symmetry.

\section{More generic quark-adjoint interactions}

We can generalize the techniques we developed for $\Tr[Q \Phi^k {\tilde Q}] $
interaction to study theories with superpotentials of the form $\Tr[Q
P(\Phi) {\tilde Q}] $, where $P(\Phi)$ is an arbitrary polynomial of finite degree.

The matrix prescription is obvious. One must compute quark contribution to
the free energy by summing all the one boundary Feynman diagrams with
arbitrary types of insertions. The vertices are given by the
monomials appearing in $P(\Phi)$, and the quark and adjoint propagators
are the inverse of their masses. 

While a formal expression for the free energy of the matrix model can be written
for a generic $P(\Phi)$, it does not seem of any particular use. Let as only 
illustrate the procedure by considering a theory with an interaction
term of the form 
\be
\Tr[Q (g_1 \Phi^{2 p_1}+ g_2 \Phi^{2 p_2}+ g_3 \Phi^{2 p_3}) {\tilde Q}] ~~,
\ee 
where we have chosen the powers of $\Phi$ to be even in
order to keep the counting easy. A general diagram containing  $k_i$
vertices of type $i$  comes with a factor of $1/( k_1 ! k_2! k_3 !)$ from the expansion of 
the exponential; the different ways of contracting the quarks give a
factor of $(k_1+k_2+k_3-1)!$; the different ways of connecting the
boundary points with non-intersecting adjoint propagators give
\cite{brezin} a factor  
\be
{(2 k_1 p_1 + 2 k_2 p_2 + 2 k_3 p_3 ) ! \over (
k_1 p_1 +  k_2 p_2 +  k_3 p_3 +1)! (k_1 p_1 + k_2 p_2 +  k_3
p_3)!}~~. 
\ee
Since each diagram contains $k_1+k_2+k_3$ quark propagators, 
$k_1 p_1 + k_2 p_2 +  k_3 p_3 $ adjoint
propagators, and $ k_1$, $k_2$, and respectively $k_3$ vertices, it is
multiplied by 
\be
{g_1^{k_1} g_2^{k_2} g_3^{k_3} \over
M_{\Phi}^{ k_1 p_1 + k_2 p_2 +  k_3 p_3}  m_{Q}^{
k_1+k_2+k_3}}~~.
\ee
Finally, the external flavor loop gives a factor of  $N_f$, while the  $(k_1 p_1 + k_2 p_2 +  k_3 p_3+1)$
color loops give the appropriate power of the glueball superfield. Thus, the 1-boundary contribution to 
the
free energy of this model is simply
\bea
&&\F_{\chi=1} = - N_f S \sum_{\stackrel{k_1,k_2,k_3 =0}{k_1+k_2+k_3 \ne0}}^{\infty}   {(k_1+k_2+k_3-1)!  \over ( k_1 ! k_2! k_3 !)}  \times \\
&&
\times  {(2 k_1 p_1 + 2 k_2 p_2 + 2
k_3 p_3 ) ! \over ( k_1 p_1 +  k_2 p_2 +  k_3 p_3 +1)! (k_1 p_1 + k_2 p_2
+  k_3 p_3)!} \left({S \over M_{\phi}}\right)^{ k_1 p_1 + k_2 p_2 +  k_3
p_3}{g_1^{k_1} g_2^{k_2} g_3^{k_3} \over m_{Q}^{ k_1+k_2+k_3}}
\nonumber
\eea
For odd monomials the above sums have to be only over combinations of
terms which give an even number of $\Phi$ propagators. It is quite easy
to see that in general such sums are hard to compute explicitly.

Nevertheless, it is possible to obtain this free energy by using the
vertex splitting procedure we used in section 3. A vertex of the form $g_n
\Tr[Q \Phi^n {\tilde Q}]$ can be thought of as arising from a theory with $n$
quarks and off diagonal Yukawa interactions 
$ g\Tr[ {\cal Q}_n \Phi{\tilde{\cal  Q}}_n{\cal P}_n]$ of the type (\ref{off1})
by integrating out the auxiliary quarks $Q_i$. 
 
Therefore, the matrix model of the theory with a polynomial
interaction $\Tr[ Q P(\Phi) {\tilde Q}] $ can be related to that of a theory with
interactions linear in $\Phi$ if one introduces $n-1$ auxiliary quarks for
each monomial $g_n \Tr[Q \Phi^n {\tilde Q}] $ coming from $P(\Phi)$. One obtains
an off-diagonal interaction matrix whose dimension is the sum of the
powers of the monomials in $P(\Phi)$ minus the number of monomials plus
one. Since the auxiliary quarks have the same mass as $Q$ \footnote{As we
explained, this does not affect integrating them out.}, one can
diagonalize the interaction matrix and obtain a theory with diagonal
Yukawa interactions $\lambda_j \Tr[Q_j \Phi {\tilde Q}_j]$, where the
$\lambda_j$ are the eigenvalues of the interaction matrix. 

As in the case discussed in section 2, the partition function
will be the sum of 1-quark regular Yukawa partition functions with
couplings  $\lambda_j$. The only difference from the equation (\ref{Kp}) 
will be that the couplings will not be proportional to the roots of unity, but will
have a more complicated form. 

To treat this theory correctly in the gauge theory, one must
perform the same steps, by integrating in the auxiliary quarks and
obtaining a theory with only linear couplings of the adjoint field $\Phi$. One can then
integrate out  $\Phi$  and obtains a theory with only quarks. For this
theory it is possible to determine that only the regular
Affleck-Dine-Seiberg superpotential is generated nonperturbatively; one
can then integrate out all the quarks and obtain a superpotential which
can be related to the matrix one. 

It appears therefore that since integrating in and out work identically on
the two sides, the equivalence of the matrix result and the gauge theory
result is ensured by the equivalence of these results for theories with
quarks with equal mass and different Yukawa couplings. This equivalence is
obvious from the computations in sections 2 and 3, and follows also from
the results of \cite{ohta} by making particular choices for the mass 
parameters and rescaling the quarks.

\section{Conclusions}

We have investigated an $U(N)$ gauge theory with adjoint and fundamental
matter
interacting via a coupling of the form $\Tr[Q \Phi^n {\tilde Q}]$. We 
have solved
the matrix model and found the exact low energy effective
superpotential. This effective superpotential is identical to that of a
theory with $n$ quarks minimally coupled to $\Phi$, with coupling
constants proportional to the $n$'th roots of unity. As expected, these
two theories are related by integrating  in/out $n-1$ quarks.

On the gauge theory side we argued that in order to determine
unambiguously the nonperturbatively generated contribution to the
superpotential one needs to first integrate in these auxiliary $n-1$ quarks, obtain a
theory with minimal couplings between the quarks and the adjoint field $\Phi$,
and then integrate out $\Phi$ to obtain a theory with $n$ massive quarks
and a quartic tree-level superpotential. One can then use standard
holomorphy and symmetry arguments to argue that the only
nonperturbative superpotential in this theory is the Affleck-Dine-Seiberg
one. By integrating out all the quarks we obtained the low energy
effective superpotential of our theory, and we found 
it agrees with the one computed in the matrix model.


We also described a method to investigate
theories with more complicated adjoint-quark couplings, of the form
$\Tr[Q P(\Phi) {\tilde Q}] $, where $P(\Phi)$ is a generic polynomial of finite
degree. We illustrated this technique by writing down the matrix free 
energy for a polynomial $P$ built out of three monomials of arbitrary even powers.
While the generalization is straightforward, the perturbation theory  is difficult to 
resume. It is also possible to further generalize this discussion by adding an arbitrary 
superpotential depending only on the the adjoint field. 

We then presented a method to solve these theories by relating them to gauge
theories with many quarks but only minimal couplings $ \lambda_{i}\Tr[
Q_i \Phi {\tilde Q}_i]$. When the polynomial contains just one monomial of order
$n$ (this is the case discussed in the first two sections of this note),
the $\lambda_{i}$ are proportional to the $n$'th roots of unity. For more
complicated polynomials, the $\lambda_{i}$ are the eigenvalues of the
quark interaction matrix.

\section*{Acknowledgments}

We would like thank Eric D'Hoker, Joe Polchinski and Per Kraus for useful discussions and support. 
The work of I.B. and S.dH. was supported in part by the NSF under Grant No. 
PHY00-99590. The work of R.R. was supported in part by the DOE under Grant No. 91ER40618 and in part by 
the NSF under Grant No. PHY00-98395. Any opinions, findings, and conclusions or recommendations 
expressed in this material are those of the authors and do not necessarily reflect the views of the 
National Science Foundation.


\begin{thebibliography}{99}


\bibitem{dv1}  
R.~Dijkgraaf and C.~Vafa,
``Matrix models, topological strings, and supersymmetric gauge theories,''
Nucl.\ Phys.\ B {\bf 644}, 3 (2002), {\tt hep-th/0206255}

\bibitem{dv2}
R.~Dijkgraaf and C.~Vafa,
``On geometry and matrix models,''
Nucl.\ Phys.\ B {\bf 644}, 21 (2002), {\tt hep-th/0207106}

\bibitem{dv3}  
R.~Dijkgraaf and C.~Vafa,
``A perturbative window into non-perturbative physics,''
{\tt hep-th/0208048}

\bibitem{ACFH}
R.~Argurio, V.~L.~Campos, G.~Ferreti and R.~Heise, ``Exact Superpotentials for Theoreis with Flavors 
via a Matrix Integral'', {\tt hep-th/0210291}

\bibitem{BERO}  
I.~Bena and R.~Roiban,
``Exact superpotentials in N = 1 theories with flavor and their matrix  model formulation,''
{\tt hep-th/0211075}



\bibitem{SUZU}
H.~Suzuki,
``Perturbative derivation of exact superpotential for meson fields from  matrix  
theories with one flavour,'' {\tt hep-th/0211052}


\bibitem{janik}  
Y.~Demasure and R.~A.~Janik,
``Effective matter superpotentials from Wishart random matrices,''
{\tt hep-th/0211082}

\bibitem{fehe}
B.~Feng, ``Seiberg Duality in Matrix Model'',
{\tt hep-th/0211202},
B.~Feng and Y.~H.~He,
``Seiberg Duality in Matrix Models II,''
{\tt hep-th/0211234}


\bibitem{naculich}
S.~G.~Naculich, H.~J.~Schnitzer and N.~Wyllard,
``The N = 2 U(N) gauge theory prepotential and periods from a  perturbative matrix model  
calculation,'' hep-th/0211123; ``Matrix model approach to the N=2 U(N) gauge theory  
with matter in the fundamental representation'', {\tt hep-th/0211254}

\bibitem{ohta}
K.~Ohta, ``Exact Mesonic Vacua from Matrix Models'', {\tt hep-th/0212025}

\bibitem{grisaru}
R.~Dijkgraaf, M.~T.~Grisaru, C.~S.~Lam, C.~Vafa and D.~Zanon,
``Perturbative computation of glueball superpotentials,''
{\tt hep-th/0211017}

\bibitem{witten}
F.~Cachazo, M.~R.Douglas, N.~Seiberg, E.~Witten, ``Chiral Rings and Anomalies in Supersymmetric Gauge 
Theory'', {\tt hep-th/0211170}

\bibitem{rrr}
I.~Bena, R.~Roiban and R.~Tatar, ``Baryons, Boundaries and Matrix Models'', {\tt hep-th/0211271 }

\bibitem{MCGR}
J.~McGreevy,
``Adding flavor to Dijkgraaf-Vafa,''
{\tt hep-th/0211009}

\bibitem{GOPA}  
R.~Gopakumar,
``N = 1 theories and a geometric master field,''
{\tt hep-th/0211100}

\bibitem{ito}
H.~Itoyama and A.~Morozov,
``The Dijkgraaf-Vafa prepotential in the context of general Seiberg-Witten theory,''
{\tt hep-th/0211245}

\bibitem{TACH}
Y.~Tachikawa,
``Derivation of the Konishi anomaly relation from Dijkgraaf-Vafa with  \\ (bi-)fundamental matters,''
{\tt hep-th/0211189}

\bibitem{Dorey}
N.~Dorey, T.~J.~Hollowood and S.~P.~Kumar,
``S-duality of the Leigh-Strassler deformation via matrix models,''
{\tt hep-th/0210239}

\bibitem{Aganagic}
M.~Aganagic, A.~Klemm, M.~Marino and C.~Vafa,
``Matrix model as a mirror of Chern-Simons theory,''
{\tt hep-th/0211098}

\bibitem{Ferrari}{
F.~Ferrari,
``Quantum parameter space and double scaling limits in N = 1 super  Yang-Mills theory,''
{\tt hep-th/0211069}
F.~Ferrari,
``On exact superpotentials in confining vacua,''
{\tt hep-th/0210135}
}

\bibitem{Fuji}{
H.~Fuji and Y.~Ookouchi,
``Comments on effective superpotentials via matrix models,''
{\tt hep-th/0210148}
}

\bibitem{DGKV}{
R.~Dijkgraaf, S.~Gukov, V.~A.~Kazakov and C.~Vafa,
``Perturbative analysis of gauged matrix models,''
{\tt hep-th/0210238}
}

\bibitem{nacu}{
S.~G.~Naculich, H.~J.~Schnitzer and N.~Wyllard,
``The N = 2 U(N) gauge theory prepotential and periods from a  perturbative matrix model  
calculation,''
{\tt hep-th/0211123}
}


\bibitem{Berenstein}{
D.~Berenstein,
``Quantum moduli spaces from matrix models,''
arXiv:hep-th/0210183.
}

\bibitem{new6}
H.~Ita, H.~Nieder, Y.~Oz, {\tt hep-th/0211261}

\bibitem{new5}
Y.~Konishi, M.~Naka, {\tt hep-th/0212020};

\bibitem{new4}
B.~Feng, {\tt hep-th/0212010};

\bibitem{new3}
H.~Itoyama, A.~Morozov, {\tt hep-th/0212032};

\bibitem{new2}
Y.~Tachikawa, {\tt hep-th/0211274};

\bibitem{new1}
R.~Dijkgraaf, A.~Neitzke and C.~Vafa, {\tt hep-th/0211194};

\bibitem{new}
R.~Argurio, V.~L.~Campos, G.~Ferretti and R.~Heise, {\tt hep-th/0211249}

\bibitem{KAPU} A.~Kapustin, 
``The  Coulomb  branch of N=1 supersymmetric gauge theory
with adjoint and fundamental matter,'' Phys.Lett. {\bf B398} (1997) 104, {\tt hep-th/9611049}

\bibitem{ADS} I.~Affleck, M.~Dine and N.~Seiberg,
``Dynamical Supersymmetry Breaking In Four-Dimensions And Its Phenomenological Implications,''
Nucl.\ Phys.\ B {\bf 256}, 557 (1985).


\bibitem{seiberg}
N.~Seiberg,
``Exact results on the space of vacua of four-dimensional SUSY gauge theories,''
Phys.\ Rev.\ D {\bf 49}, 6857 (1994), {\tt hep-th/9402044}


\bibitem{brezin}
E.~Br\'{e}zin, C.~Itzykson, G.~Parisi and J.~B.~Zuber, ``Planar Diagrams'', Comm. Math. Phys. 59 (1978), 
35-51

\end{thebibliography}
\end{document}